# INHOMOGENEOUS COSMOLOGICAL MODEL IN LYRA GEOMETRY

F.Rahaman, S Chakraborty and J.Bera

Dept. of Maths., Jadavpur University, Kolkata – 700032, India

E.Mail: farook_rahaman@yahoo.com

**Abstract:**
Exact solutions are obtained for an inhomogeneous cosmological model in normal gauge for Lyra's geometry. Some properties of the model have also been discussed.

## Introduction:

The origin of structure in the Universe is one of the greatest cosmological mysteries even today. So the aim of cosmology is to determine the large-scale structure of the physical Universe. Since the discovery of general relativity by Einstein there have been numerous modification of it. Long ago, since 1951, Lyra [G. Lyra (1951)] had been considered an alternating theory of Einstein gravity. He suggested a modification of Riemannian geometry by introducing a gauge function into the structureless manifold that bears a close resemblance to Weyl's geometry. In consecutive investigations, Sen,[D.K.Sen (1957)] Sen and Dunn [D.K.Sen & K.A.Dunn (1971)] proposed a new scalar-tensor Theory of gravitation and constructed an analog of the Einstein field equations

based on Lyra's geometry which is normal gauge may be written as

$$R_{ik} - \frac{1}{2} g_{ik} R + \frac{3}{2} \phi_i \phi_k - \frac{3}{4} g_{ik} \phi_j \phi^j = -\chi T_{ik} \tag{1}$$

where $\phi_i$ is the displacement vector and other symbols have their usual meaning as in Riemannian geometry. After that so many works were done by several authors in scalar tensor theory and cosmology within the framework of Lyra's geometry [K.S.Bharma (1974); S.B.K.Shetti and B.B.Waghmode,(1982); A.Beesham (1986); H.D.Soleng (1987); T.Singh and G.F.Singh (1991); G.P.Singh and K. Desikan (1997); F.Rahaman and J.Bera (2001)]. As far our knowledge all the above works related to spatially homogeneous models. The observational fact that the Universe is not exactly homogeneous and isotropic. Therefore we have thought it worthwhile to study inhomogeneous cosmological model on Lyra's geometry.

## 2. Field Equations:

The time like displacement vector $\phi_i$ in (1) is given by

$$\phi_i = (\beta(z,t),0,0,0) \tag{2}$$

We take a perfect fluid from the energy momentum tensor:

$$T_{ik} = (p+\rho) W_i W_k - p g_{ik} \tag{3}$$

together with commoving co-ordinates $W^i W_i = 1$.

We consider a plane symmetric space time

$$ds^2 = e^A(dt^2 - dz^2) - e^C(dx^2 + dy^2) \tag{4}$$

where A, C are functions of z and t.

With Eqs. (2) - (4), the field Eqs. (1) become

$$\frac{e^{-A}}{4}(-4C'' - 3C'^2 + 2A'C') + \frac{e^{-A}}{4}(\dot{C}^2 + 2\dot{A}\dot{C}) - \frac{3}{4}\beta^2 e^{-A} = \chi\rho \tag{5}$$

$$\frac{e^{-A}}{4}(-C'^2 - 2C'A') + \frac{e^{-A}}{4}(4\ddot{C} + 3\dot{C}^2 - 2\dot{A}\dot{C}) + \frac{3}{4}e^{-A}\beta^2 = -\chi p, \tag{6}$$

$$\frac{e^{-A}}{4}(-2A'' - 2C'' - C'^2) + \frac{e^{-A}}{4}(2\ddot{A} + 2\ddot{C} + \dot{C}^2) + \frac{3}{4}\beta^2 e^{-A} = -\chi p, \tag{7}$$

$$\frac{1}{2}(\dot{C}' + \dot{C}(A' - C') + \dot{A}C') = 0, \tag{8}$$

Further the conservation equations $T^{ik}{}_{;k} = 0$ reduce to

$$\chi\dot{\rho} + \frac{3}{2}\beta\dot{\beta} + 2\left[\chi(p + \rho) + \frac{3}{2}\beta^2\right][\dot{A} + 2\dot{C}] = 0 \tag{9}$$

and

$$\chi p' + \frac{3}{2}\beta\beta' + 2\left[\chi(p + \rho) + \frac{3}{2}\beta^2\right][A' + 2C'] = 0 \tag{10}$$

Here prime and dot are differentiations w. r. t. z and t respectively, we also assume the equation of state

$$p = \lambda\rho, \qquad 0 \leq \lambda \leq 1, \tag{11}$$

## 3. Solutions:

To solve the field equations we shall assume the separable form of the metric coefficients as follows:

$$A = A_1(z) + A_2(t); \qquad C = C_1(z) + C_2(t). \qquad (12)$$

Now from (8), by using (12), we get

$$A_1 = mC_1, \qquad (13)$$

$$A_2 = (1-m)C_2, \qquad (14)$$

where *m* is the separation constant.

Subtracting (7) from (6) and using (13) and (14), we get

$$(2m+2)C_1'' - 2mC_1'^2 = -2m\ddot{C}_2 - 2m\dot{C}_2^2 = -n \qquad (15)$$

where *n* is separation constant.

From (15) we get

$$C_1 = \frac{1}{b}\ln\sinh(abz), \qquad (16)$$

$$C_2 = \ln\cosh(at), \qquad (17)$$

where $a^2 = \dfrac{n}{2m}, \; b = \dfrac{m}{m+1}.$

We also get following expressions of $\rho$, $p$ and $\beta^2$ as follows:

$$\rho = \frac{-(b+1)}{(1-\lambda)\chi} \operatorname{cosech}^{\frac{1}{b}+2}(abz)\operatorname{sech}(at), \qquad (18)$$

$$p = \frac{-(b+1)\lambda}{(1-\lambda)\chi} \operatorname{cosech}^{\frac{1}{b}+2}(abz)\operatorname{sech}(at), \qquad (19)$$

$$\beta^2 = \frac{1}{3}\left\{\frac{4(\lambda b+1-a^2)}{1-\lambda} - \frac{ba^2}{2} + (3-2m)a^2[\tanh^2(at) - \coth^2(abz)]\right\} \qquad (20)$$

The kinematic parameters expansion $\theta$, shear $\sigma^2$ and acceleration $\dot{V}_z$ are given by

$$\theta = \frac{1}{2}\operatorname{cosech}^{1/2b}(abz)\operatorname{sech}^{1/2}(at)(4-2m)a\tanh(at), \qquad (21)$$

$$\sigma^2 = \frac{1}{6}\operatorname{cosech}^{1/b}(abz)\operatorname{sech}(at)(1-2m)a^2\tanh^2(at), \qquad (22)$$

$$\dot{V}_z = \frac{-m}{2}a\coth(abz). \qquad (23)$$

## 4. Discussion:

We have obtained an inhomogeneous cosmological model on Lyra geometry. We find the metric as

$$ds^2 = \sinh^{m/b}(abz)\cosh^{(1-m)}(at)(dt^2 - dz^2) - \sinh^{1/b}(abz)\cosh(at)(dx^2 + dy^2) \quad (24)$$

We note that our space time is singularity free. The energy condition imply that $\rho > 0$. So we must have $(b+1) < 0$ since $\lambda \leq 1$. Hence the arbitrary constant $m$ is confined between

$$-1 < m < -\frac{1}{2}.$$

The space time of the class of solutions is to represent an expanding Universe for $t > 0$. The energy density and pressure tend to zero as time, increases indefinitely. Also these variables tend to zero as Z increases indefinitely. The kinematical variables $\theta$ and $\sigma^2$ will be vanished as $t \to \infty$ but diverge as $Z \to \infty$. We also see that there is no possibility that the models may get isotropized at some later i.e. it remains anisotropy for all times. For $b = -1$ i.e. $m = -(1/2)$, $\rho = p = 0$ and this class of solutions degenerates into singularity – free vacuum solutions based on Lyra geometry. It is interesting to note that the displacement vector i.e. the concept of Lyra geometry still exists after infinite times.

Acknowledgement:
                    We are thankful to the members of Relativity and Cosmology center, Jadavpur University for helpful discussions.